\documentclass[aps,prb,twocolumn,showpacs]{revtex4-1}

\usepackage{amsfonts,amssymb,amsmath}
\usepackage{mathtools}
\usepackage[]{graphics,graphicx,epsfig}
\usepackage{amsthm,multirow}
\usepackage{color}
\begin{document}

\title{Ultrabright fiber-coupled ploarization-entangled photon source with spectral brightness surpassing 2.0 MHz/mW/nm}

\author{Kyungdeuk \surname{Park}}
\email{kpark@add.re.kr}

\author{Jungmo Lee}

\author{Dong-Gil Im}

\author{Dongkyu Kim}

\author{Yong Sup \surname{Ihn}}
\email{yong0862@add.re.kr}

\affiliation{Emerging Science and Technology Directorate, Agency for Defense Development, Daejeon 34186, Korea}

\date{\today}

\begin{abstract}
We present an ultrabright polarization-entangled photon source that is optimally coupled into single-mode fibers (SMFs).
This study theoretically and experimentally examines the characteristics of spontaneous parametric down-conversion (SPDC) photons, including their spectrum, bandwidth, emission angle, and intensity, as functions of crystal length, temperature and beam waist condition.
Notably, we measure the collinear spatial modes of photon-pairs and collection optics under various beam waist conditions and analyze them using a collinear Gaussian approximation model.
By employing a simple mode-matching optical setup, we optimize the SMF coupling and heralding efficiencies of the photon-pairs.
Consequently, we achieve a spectral brightness exceeding 2.0 MHz/mW/nm from a fiber-coupled entangled photon source, utilizing a 30-mm ppKTP crystal inside a polarization Sagnac interferometer. 
This represents the highest spectral brightness of SPDC photons generated using a CW laser pumped bulk crystal to date.
Polarization entanglement was verified by a quantum state tomography and a polarization-correlation measurement.
The fidelity of the entangled state is measured to be 97.8 $\%$ and the Bell-CHSH value $\it S$ $=$ 2.782 $\pm$ 0.04.
The results obtained here provide practical insights for designing high-performance SPDC sources for satellite-based communication and long-distance optical links with extremely high-photon loss.
\end{abstract}

\maketitle

\section{Introduction}

Entangled photon sources with sufficient brightness and entanglement visibility are crucial for realizing global quantum networks \cite{Nature08Kimble,NP16Vlaivarthi}.
These sources are essential for interconnecting satellites, ground stations, and inter-satellite links \cite{Science17Yin,Nature17Liao,RMP22Lu}, even under extremely high-loss conditions \cite{PRL18Cao}.
Various novel methods using different media have been proposed for the efficient generation of photon-pairs \cite{ACSphoton21Zeuner,PRXquantum21Steiner,QST23Bruns,arXiv24Shiu}, but despite the low parametric gain, entangled photon sources based on SPDC process in $\chi^{(2)}$ nonlinear crystals remain highly regarded for satellite-based quantum communications due to their high purity and entanglement fidelity at high generation rates.

Efforts to enhance the brightness of entangled photon sources have explored various nonlinear crystals in both bulk and waveguide structures \cite{PRL88Shih,PRL95Kwiat,OE07Fujii,OE09Chen}, different phase-matching conditions including type-0 \cite{OE12Steinlechner,JOSAB14Steinlechner,SR17Jabir}, type-I \cite{OE08Fiorentino}, and type-II \cite{JOSAB14Steinlechner,PRA05Lee,OE11Halevy,OE13Jeong}, and a range of experimental schemes \cite{OE13Steinlechner,APL20Lohrmann,QST21Lee}.
In the field of satellite-based long-distance quantum communication, both bright quantum light sources and fiber-coupling technology that minimizes photon-pair loss are required to effectively transmit and receive quantum states.
The beam waist of photon-pairs generated in the crystal varies with the pump waist, significantly impacting the optical system design for SMF coupling.
Both theoretical \cite{PRA05Ljunggren,PRA10Bennink} and experimental \cite{OE12Steinlechner,SR17Jabir,OE07Fedrizzi,OE13Guerreiro} studies have been conducted to optimize the pump waist focused at the crystal center and the associated collection optics parameters.
High coupling efficiency of fiber-coupled entangled photon sources has also been reported \cite{OE12Steinlechner,JOSAB14Steinlechner,SR17Jabir}.

In this work, we directly measure the spatial mode profiles of photon-pairs generated in type-0 ppKTP crystals and compare them with the fundamental mode profile created by a SMF and optical fiber-collimator.
Based on these measurements, we employ a simple optical mode-matching system to optimize the collinear Gaussian mode of the propagated SPDC photons to match as closely as possible with the fundamental mode of the collection optics. 
As a result, we achieve a spectral brightness as high as 2.0 MHz/mW/nm for SMF coupling, more than almost twice the highest value reported to date.
These results provide a simple and practical method for designing fiber-coupled, highly efficient entangled photon sources needed for long-distance free-space quantum communication in environments with significant photon loss. 

The paper is organized as follows: in section 2, we investigate the influence of the crystal length, temperature and pump waist on the characteristics of SPDC photons, including photon-pair spectrum, bandwidth, emission angles, and intensities.
In section 3, we identify the optimal spatial mode-matching between the SPDC photon-pairs and the SMF to achieve effective coupling.
In section 4, we describe the ultrabright source of polarization-entangled photons and present the results obtained in spectral brightness and quality of entanglement.

\section{SPDC photon-pair characteristics in type-0 ppKTP crystal}

The SPDC process in second-order nonlinear crystals generates pairs of lower-frequency photons, known as signal and idler photons, from a pump photon.
SPDC can manifest in different types based on the polarizations of the interacting fields: type-0, type-I, and type-II.
In type-0 SPDC, unlike type-I and type-II processes, the pump and SPDC photons share the same polarization.
SPDC is feasible under conditions that satisfy both energy and momentum conservation.
Recent advancements in periodic poling technology have enabled quasi-phase-matching for a wide range of wavelengths and polarizations.

\begin{figure}[t]
\centering
\includegraphics[width=0.47\textwidth]{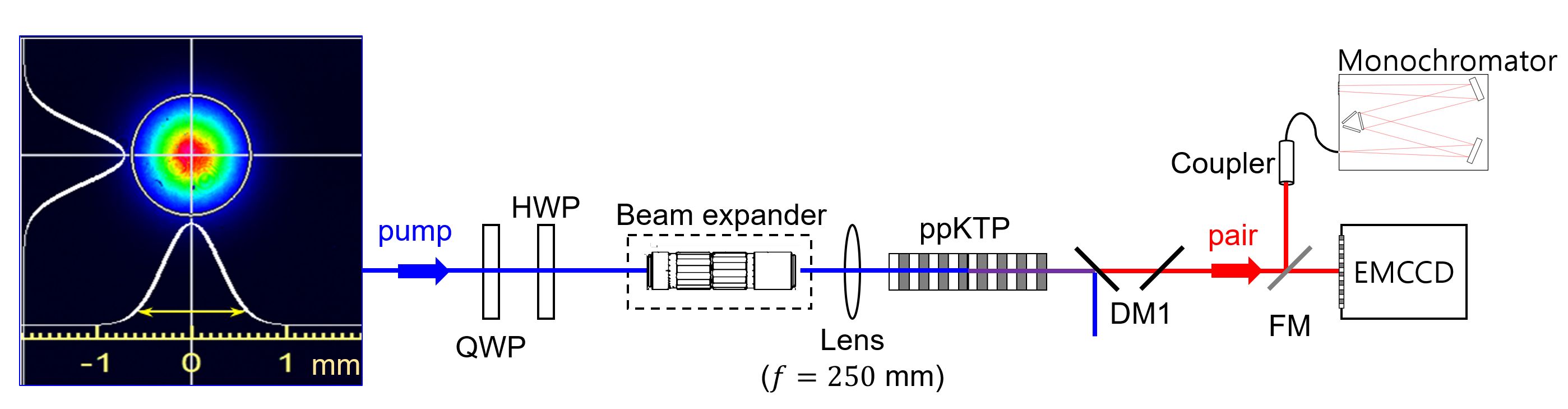}
\caption{Experimental schematic for measuring the characteristics of photon-pairs generated via collinear type-0 SPDC. QWP; quarter-wave plate, HWP; half-wave plate, DM1; dichroic mirror 1, FM; flip mirror, EMCCD; electron-multiplying charge-coupled device.}
\label{fig1}
\end{figure}

From the interaction Hamiltonian, the two-photon state generated via SPDC can be described as
\begin{equation}
\begin{split}
\vert\psi(\omega_{s},\omega_{i})\rangle \propto &  \int\int d\omega_{s}d\omega_{i}\xi_{p}(\omega_{s}+\omega_{i})\text{sinc}\left( \frac{\Delta k L}{2} \right) \\
& \times \hat{a}_{s}^{\dagger}(\omega_{s})\hat{a}_{i}^{\dagger}(\omega_{i})\vert\ \text{vac} \rangle.
\label{eq1}
\end{split}
\end{equation}
Here, $\xi_{p}$ denotes the pump spectral envelop function, and $\Delta k(\theta,\omega)=k_{p}(\theta_{p},\omega_{p})-k_{s}(\theta_{s},\omega_{s})-k_{i}(\theta_{i},\omega_{i})-\frac{2\pi}{\Lambda}$ represents the phase mismatch.
$L$ and $\Lambda$ are the crystal length and the poling period of the crystal, respectively, and $\omega_{p,s,i}$ and $k_{p,s,i}$ denote the angular frequencies and wave vectors of the pump, signal, and idler photons, respectively.
The wavevector $k$ is a function of the refractive index of the crystal $n(\lambda,T)$, which depends on the crystal's temperature $T$ and the wavelength $\lambda(=2\pi c/\omega)$ of the interacting field inside the crystal.
The intensity (or number of photons) of two-photon state can be expressed as $P_{si}=\int\int d\omega_{s}d\omega_{i}\left|\psi(\omega_{s},\omega_{i})\right|^{2}$.

Figure \ref{fig1} illustrates the experimental schematic employed to investigate the characteristics of the SPDC photon source.
In a non-degenerate, collinear type-0 $(e\to e+e)$ phase-matching configuration, photon-pairs are generated within a 30-mm long ppKTP crystal (Raicol) with a poling period of $\Lambda=$3.425 $\mu$m and a nonlinearity $d_{eff}=\frac{2}{\pi}d_{33}=12$ pm/V \cite{CRSSM99Satyanarayn}.
The crystal is pumped by a continuous-wave, single-frequency diode laser (Topmode 405HP, Toptica) operating at a wavelength of 405.143 nm.
The pump waist (1/$e^{2}$ radius) measures approximately 625 $\mu$m as depicted in Fig. \ref{fig1}, and polarization control is achieved using both a QWP and a HWP.

To measure the photon-pair intensity relative to the pump waist, the waist size focused at the crystal center is adjusted using a variable beam expander (ZBE1A, Thorlabs) and lens.
Photon-pairs around the 810 nm wavelength, generated through the SPDC process, pass through two dichroic mirrors (DMLP650, Thorlabs) before entering the detection setup, where the residual pump beam is filtered out.
Initially, utilizing a monochromator (SR-5001-B1, Andor tech.) and an EMCCD (Ultra888, iXon), we investigate the influence of crystal temperature and length on the spectra, spatial modes, and brightness of the SPDC source.

\begin{figure}[t]
\centering
\includegraphics[width=1\linewidth]{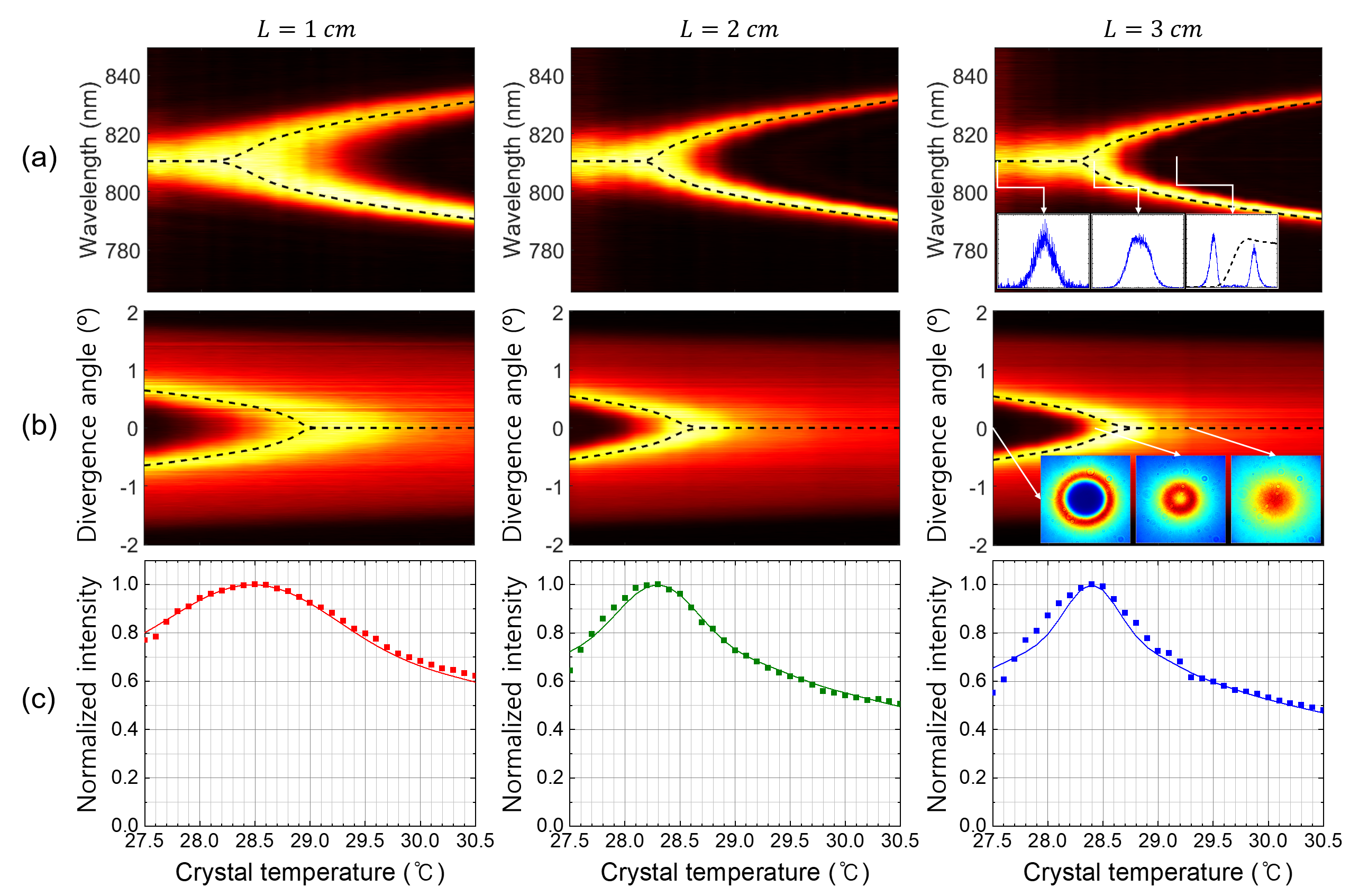}
\caption{(a) Measured spectra, (b) emission angles, and (c) normalized intensities of SPDC photons as a function of crystal temperature. Each column presents the corresponding data for three different lengths: 1, 2, and 3 cm. The dashed black lines and colored solid lines represent theoretical curves derived from numerical calculations utilizing Eq. \ref{eq1}. Insets in the third column of (a) and (b) display measured spectra and spatial modes of photon-pairs generated from the 3-cm-long crystal at 27.5, 28.4, and 29.3 $^\circ$C. The black dashed line in the inset of 29.3 $^\circ$C spectrum represents the transmission of dichroic mirror 2 (DM2) as shown in Fig. \ref{fig4}(a).}
\label{fig2}
\end{figure}

Figure \ref{fig2} presents the measured characteristics of photon-pairs generated via collinear type-0 SPDC in ppKTP crystals.
Figure \ref{fig2}(a) shows the phase-matching spectra as a function of crystal temperature for different crystal lengths. 
In this measurement process, photon-pairs emitted from ppKTP are filtered using DM1 and coupled into a SMF, followed by spectral measurement using a monochromator and an EMCCD.
By connecting the spectra of photon-pairs at varying temperatures, phase-matching spectra corresponding to the temperature variation are obtained.
As the crystal temperature increases, the photon-pair spectrum transitions from a broad Gaussian shape with a large full-width at half-maximum (FWHM) bandwidth to separate peaks for the signal and idler spectra, accompanied by a reduction in their respective bandwidth.
The spectral changes due to temperature variation appear independent of the crystal length; however, increasing crystal length decreases overall FWHM bandwidth of the spectrum.
The black dashed lines indicate wavelengths where brightness is maximized with respect to temperature.
Figure \ref{fig2}(b) displays the divergence angles of SPDC photons as a function of crystal temperature for different crystal lengths, obtained from spatial images captured directly by the EMCCD positioned approximately 18 cm from the crystal center.
At lower temperature, SPDC photons exhibit a ring-shaped spatial distribution, whereas above 28.5 $^\circ$C, photon-pairs are predominantly generated in a collinear direction.
The black dashed lines represent divergence angles where brightness is maximized with respect to temperature.

Figure \ref{fig2}(c) presents the normalized intensities of SPDC photons obtained from the spatial images, calculated by summing photon counts across all pixels of the EMCCD sensor.
At peak intensities, the spatial mode exhibits a doughnut-like shape rather than a perfect Gaussian, indicating closely positioned wavelengths for the signal and idler photons.
Thus, optimizing the crystal temperature is necessary to achieve wavelength separation and high photon-pair generation rates, as generation rates decrease with increasing temperature.
For complete separation of signal and idler photons, a temperature of 29.3 $^\circ$C is set, and wavelength separation is accomplished using a DM2 (T810lpxr, Chroma).
The center wavelengths of the signal and idler are 796 and 824 nm, respectively, with similar FWHM bandwidths of approximately 4.9 nm.

To observe variations in the photon-pair generation rate based on the pump waist focused at the crystal center, we first directly measure the intensity of the generated photon-pairs without SMF coupling using an EMCCD.
Similar to the earlier intensity measurements conducted with temperature variations, the intensity for each pump waist is obtained by summing the counts measured at all pixels.
To adjust the size of the incident pump waist, we employ a variable beam expander with a magnification range of 0.6$\times$ to 2.5$\times$, positioned in front of the pump focusing lens ($f=250$ mm), as shown in Fig. \ref{fig1}.
The waist of a Gaussian beam is defined as the radius where the irradiance reaches $1/e^{2}$ of its maximum value.
The waist of the pump beam focused at the crystal center can be calculated using the equation, $w_{0}=\frac{M^{2}\lambda f}{\pi w_{in}}$, where $w_{in}$ represents the pump waist incident on the lens, $M^{2}$ is the beam quality factor (assumed to be 1), $\lambda$ is wavelength, and $f$ is the focal length of lens.
In Fig. \ref{fig1}, a pump beam with a waist of 625 $\mu$m passes through a beam expander, allowing $w_{in}$ to be adjusted from 312 to 1630 $\mu$m depending on the magnification.
Consequently, $w_{0}$ at the crystal center ranges from 91 to 22 $\mu$m.

\begin{figure}[t]
\centering
\includegraphics[width=1\linewidth]{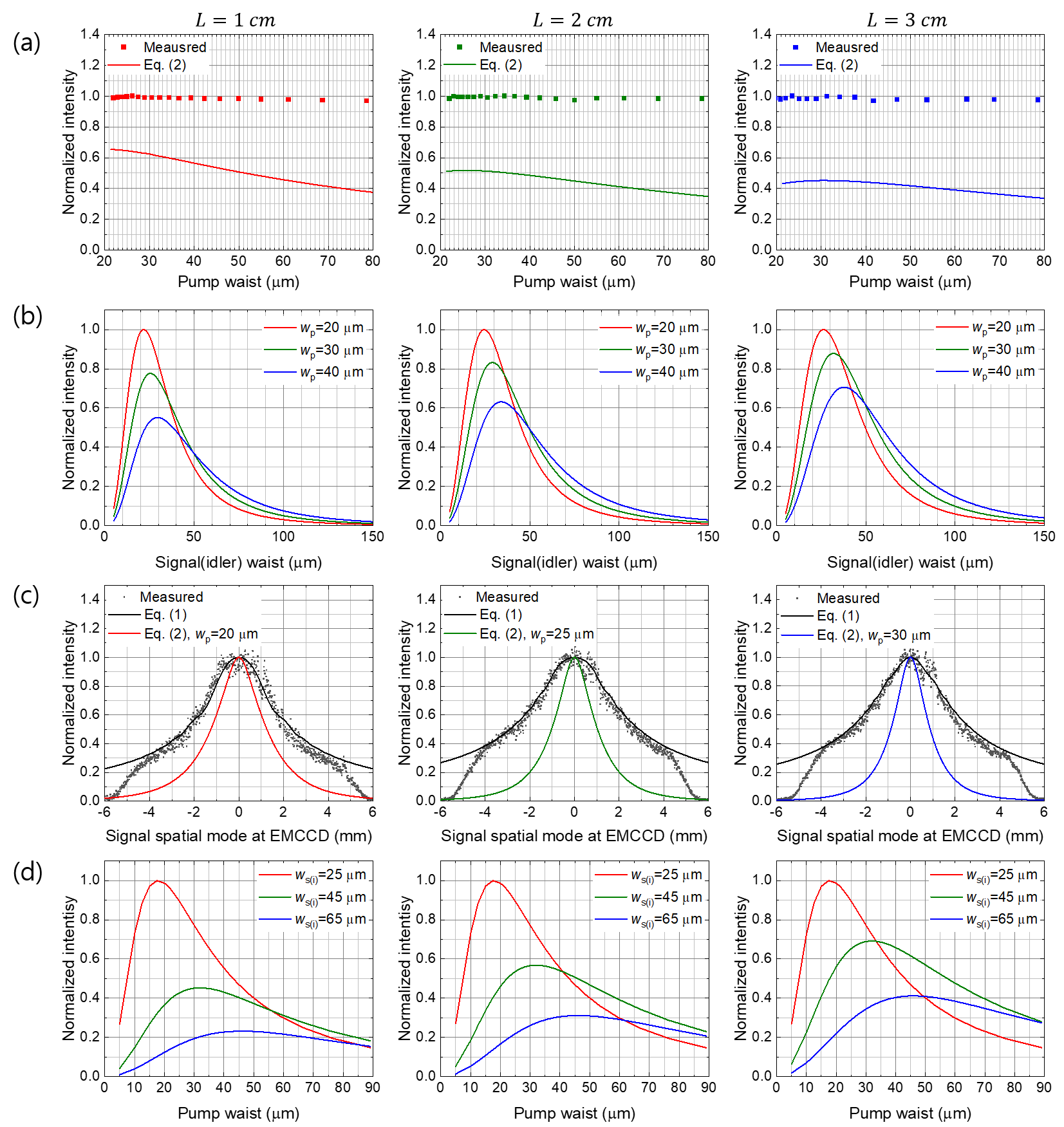}
\caption{(a) Normalized intensities of SPDC photons as a function of pump waist for different crystal lengths at 29.3 $^\circ$C. The colored solid lines represent theoretical curves based on the Eq. (\ref{eq2}). (b) Numerically calculated normalized intensities as a function of signal (idler) waist for varying pump waists. (c) Spatial mode profiles of SPDC photons captured by EMCCD. The black and colored curves are derived using Eq. (\ref{eq1}) and (\ref{eq2}), respectively. (d) Numerically calculated normalized intensities as a function of pump waist for fixed signal (idler) waists.}
\label{fig3}
\end{figure}

Figure \ref{fig3}(a) presents normalized data acquired through the direct measurements by the EMCCD, depicting the intensity of photon-pairs generated from the crystals of different lengths in response to varying the pump waist.
The intensity of photon-pairs remains constant regardless of the changes in the pump waist, which can be regarded as the measurement result of all SPDC photon modes generated in the crystal by the pump beam.
However, in SMF coupling schemes, it is necessary to consider only the collinear Gaussian modes among all generated modes.
To implement high-efficiency fiber-coupled quantum light sources, the collection optics must ensure that the modes of SPDC photons match well with the fundamental mode of the SMF.
In such schemes, the two-photon state can be represented by an electric field operator containing collinear Gaussian modes for each frequency component as $\hat{E}^{\dagger}\propto\int d\omega\frac{w}{q}\text{exp}\left(-\frac{x^{2}+y^{2}}{q}+ikz\right)\text{exp}(-i\omega t)\hat{a}(\omega, t)$, where $q=w^{2}+\frac{2iz}{k}$ \cite{PRA10Bennink}.
The two-photon state can be expressed by
\begin{equation}
\vert\psi(\omega_{s},\omega_{i})\rangle \propto \int\int d\omega_{s}d\omega_{i}\phi(\omega_{s}+\omega_{i})\hat{a}_{s}^{\dagger}(\omega_{s})\hat{a}_{i}^{\dagger}(\omega_{i})\vert\ \text{vac} \rangle,
\label{eq2}
\end{equation}
where
\begin{equation}
\phi(\omega_{s}+\omega_{i}) \propto \frac{w_{p}w_{s}w_{i}}{\sqrt{\lambda_{p}\lambda_{s}\lambda_{i}}}\int_{-L/2}^{L/2} \frac{\text{exp}[i\Delta\Phi z]}{q_{s}^{*}q_{i}^{*}+q_{p}q_{i}^{*}+q_{p}q_{s}^{*}}dz.
\label{eq3}
\end{equation}
Here, $\Delta \Phi = k_{p}-k_{s}-k_{i}-\frac{2\pi}{\Lambda}$ and $q_{a}=w_{a}^{2}+\frac{2z}{k_{a}}i$.
The subscript $a\in \{ p,s,i \}$ denotes the pump, signal, and idler.
After substituting the equation for $q_{a}$ into Eq. (\ref{eq3}) and simplifying, it becomes as follows:
\begin{equation}
\begin{split}
\phi(\omega_{s}+\omega_{i}) \propto & \frac{w_{p}w_{s}w_{i}}{\sqrt{\lambda_{p}\lambda_{s}\lambda_{i}}} \int_{0}^{L/2} \frac{\left(-Az^{2}+C\right)\text{cos}\left(\Delta\Phi z\right)}{\left(-Az^{2}+C\right)^{2}+\left(Bz\right)^{2}} \\
& -\frac{Bz\text{sin}\left(\Delta\Phi z\right)}{\left(-Az^{2}+C\right)^{2}+\left(Bz\right)^{2}}dz,
\label{eq4}
\end{split}
\end{equation}
where $A=4\left(\frac{k_{p}-k_{s}-k_{i}}{k_{p}k_{s}k_{i}}\right)$, $B=2\left(\frac{w_{p}^{2}+w_{s}^{2}}{k_{i}}-\frac{w_{s}^{2}+w_{i}^{2}}{k_{p}}-\frac{w_{i}^{2}+w_{p}^{2}}{k_{s}}\right)$, and $C=w_{p}^{2}w_{s}^{2}+w_{s}^{2}w_{i}^{2}+w_{i}^{2}w_{p}^{2}$.
Finally, the intensity of two-photon state can be calculated as $P_{si}=\int\int d\omega_{s}d\omega_{i}\left|\psi(\omega_{s},\omega_{i})\right|^{2}$.
\begin{figure}[t]
\centering
\includegraphics[width=1\linewidth]{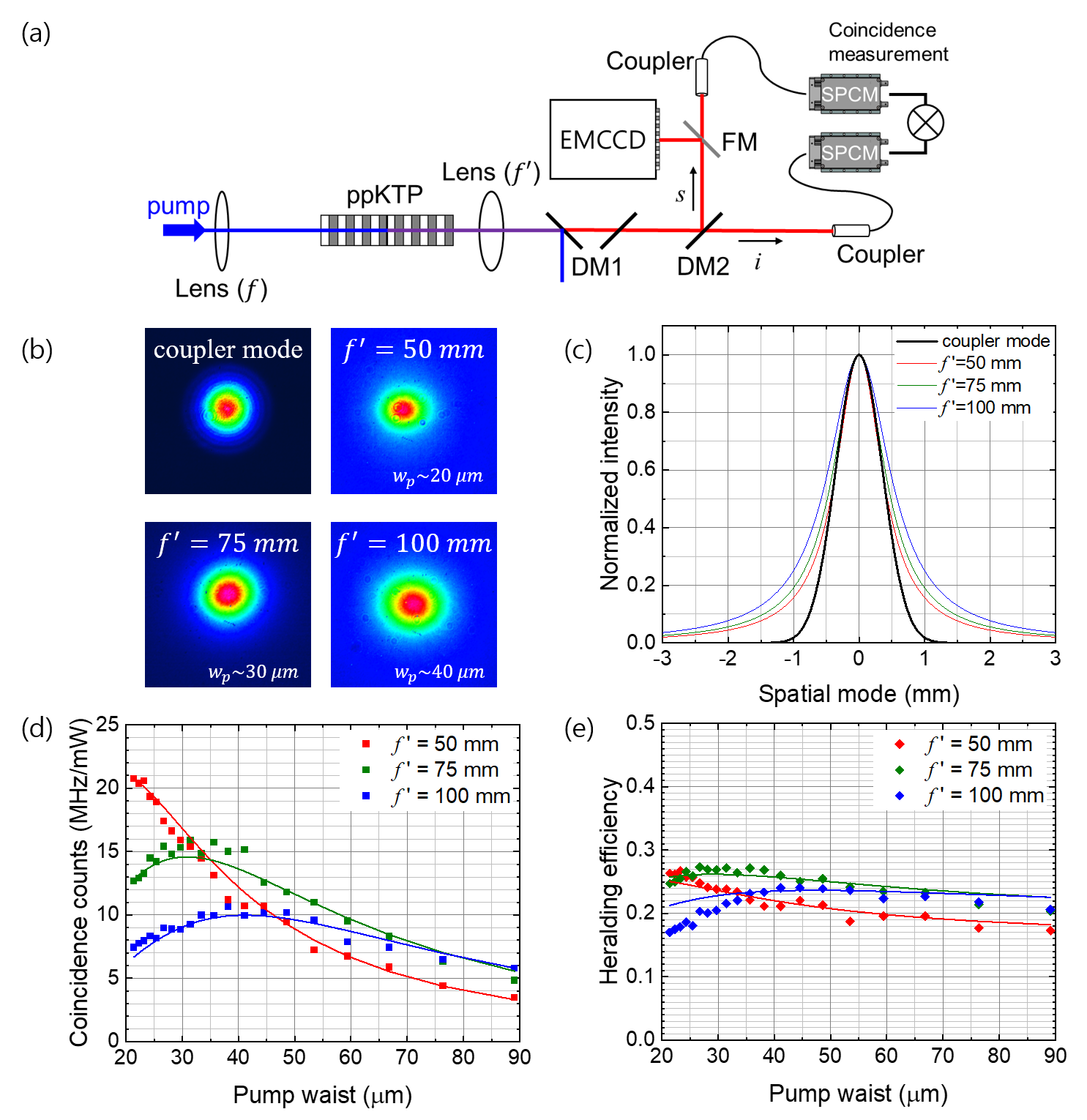}
\caption{(a) Spatial mode-matching setup for efficient optical coupling. (b) Gaussian mode images of the fiber-coupler and spatial mode images of signal photons propagated under different collimation conditions ($f'=$50, 75, and 100 mm) for each pump waist ($w_{p}=$20, 30, and 40 $\mu$m). (c) Fit-curves of spatial modes obtained from Fig. \ref{fig4}(b). (d) Coincidence counts as a function of pump waist for different collimations. (e) Heralding efficiencies as a function of pump waist for different collimations. The colored lines denote theoretical curves based on Eq. (\ref{eq2}). DM2; dichroic mirror for signal/idler separation, SPCM; single photon counting module, TCSPC; time-correlated single counting module.}
\label{fig4}
\end{figure}
The solid lines in Fig. \ref{fig3}(a) represent theoretical curves obtained using Eq. (\ref{eq2}), normalized based on the ratio of the area under the measured curve to the area calculated, as depicted in Fig. \ref{fig3}(c), according to the pump waist.
The significant discrepancy between the measured data and theoretical values arises because the theoretical calculations consider only collinear Gaussian modes, while the measured data includes collinear Gaussian modes as well as non-collinear Gaussian and non-Gaussian modes.
Theoretical values exhibit a gradual decrease as the pump waist decreases or increases, indicating that if the pump focus is too weak or too strong, the proportion of non-collinear and non-Gaussian modes increases due to the presence of multi-mode emissions in the transverse and longitudinal directions \cite{PRA05Ljunggren}.

Figure \ref{fig3}(b) presents numerically calculated intensities, normalized as a function of the signal (idler) waist, generated under different pump waists.
The values of pump waists are consistent with the experimental scheme using mode-matching between the signal (idler) and the collection optics.
The signal and idler are assumed to have the same beam waist given their small wavelength difference.
For a fixed pump waist, the signal (idler) is generated with a wide spatial distribution, and the photon-pair intensity depending on the pump waist follows a different distribution according to the signal (idler) waist.
As the pump beam tightens its focus, the signal (idler) waist distribution becomes narrows, leading to an increase in the maximum intensity.
Additionally, the signal (idler) waist distributions become progressively broader for the same pump waist as the crystal length increases.

Figure \ref{fig3}(c) displays the spatial mode profiles of SPDC photons measured by an EMCCD located 18 cm behind the ppKTP crystals.
The abrupt drop at both ends of the measured data is attributed to reduced sensitivity at the edges of the EMCCD sensor, occurring because the mode of the incoming photon-pairs is larger than the sensor's active area.
The black solid lines represent theoretical curves obtained using Eq. (\ref{eq1}), taking into account the emission angle of the photon-pairs.
The colored lines depict theoretical curves calculated using Eq. (\ref{eq2}), corresponding to pump waists at 20, 25, and 30 $\mu$m for each length yielding the maximum intensities, as shown in Fig. \ref{fig3}(a).
Comparison with the measured data indicates that the proportion of collinear Gaussian photon-pairs is relatively small compared to all modes of photon-pairs emitted from the crystal at specific pump waists.
In the SMF coupling scheme, primarily the collinear Gaussian modes of the signal and idler, well-matching the modes generated by the SMF and fiber-coupler, are coupled.
The spatial mode profile exhibits a Lorentzian distribution rather than a Gaussian distribution, arising from the superposition of collinear Gaussian modes with broad distributions in signal (idler) waist corresponding to each pump waist.
Furthermore, these spatial mode profiles become narrower as the crystal length increases.

Theoretical normalized intensities as a function of pump waist for different signal (idler) waists are shown in Fig. \ref{fig3}(d).
The tendency of collinear Gaussian mode intensities is depicted with respect to the crystal lengths and the signal (idler) waists.
Once the fundamental mode of the fiber-coupler is determined, conditions for obtaining signal (idler) modes closest to this mode can be achieved through proper optics configuration, and the SPDC photon intensity can be determined according to changes in pump waists.
In section 3, the SPDC modes generated under different pump conditions are directly compared with the mode generated by a fiber-coupler, aiming to identify conditions that yield maximum coupling efficiency with the collection optics setup established in this work.

\section{Spatial mode-matching for efficient optical coupling of SPDC photons}
In this section, we examine the optimal conditions for coupling photon-pairs into SMFs.
Efficient coupling necessitates a good match between the spatial modes of the signal (idler) photons and the fundamental mode of the SMF.
We directly compare the mode profiles of the emitted photon-pairs with the fiber-coupler modes to identify the experimental conditions that maximize coupling efficiency.

The experimental setup, depicted in Fig. \ref{fig4}(a), involves maintaining the temperature of the 3-cm-long crystal at $T=$ 29.3 $^\circ$C to fully separate the signal and idler wavelengths using a dichroic mirror (DM2).
The initial pump beam, with $w_{p}=$ 625 $\mu$m, is resized and focused at the center of the crystal by passing through a beam expander and a lens with a focal length of 250 mm.
The collection optics comprises a fiber collimator (60FC-4-M8-10, Sch$\ddot{a}$fter+Kirchhoff) with a focal length of 8.1 mm and an SMF (780HP, Thorlabs) with a mode-field radius of 2.5 $\mu$m.
The photon-pairs are collimated by a lens with a focal length $f'$ and separated into signal and idler photons by DM2, while the residual pump beam is filtered out by DM1.
To measure the spatial mode of the signal photons coupled into the collection optics, a flipper mirror is placed in front of the fiber collimator, and the signal photons are monitored with an EMCCD.
The mode generated by the collection optics is directly measured by the EMCCD using the back-coupled 810 nm laser beam exiting the fiber-coupler.

Figure \ref{fig4}(b) displays the fiber-coupler mode image and the signal photon mode images under different beam collimations ($f'=$50, 75, and 100 mm) corresponding to each pump waist ($w_{p}=$20, 30, and 40 $\mu$m), measured by EMCCD.
From the obtained images, spatial mode profile data along the x-axis are extracted, and curves fitted to the four mode images are shown in Fig. \ref{fig4}(c).
The black fit-line represents the fiber-coupler mode, indicating a Gaussian distribution, and its waist is approximately 868 $\mu$m.
The colored fit-lines show the spatial modes of signal photons for each $w_{p}$ and $f'$.
The measured signal photon modes follow a Lorentzian distribution rather than a Gaussian one due to the overlap of signal photons generated with a wide distribution of beam waists for a fixed pump waist (see Fig. \ref{fig3}(b)).
\begin{figure}[t]
\centering
\includegraphics[width=1\linewidth]{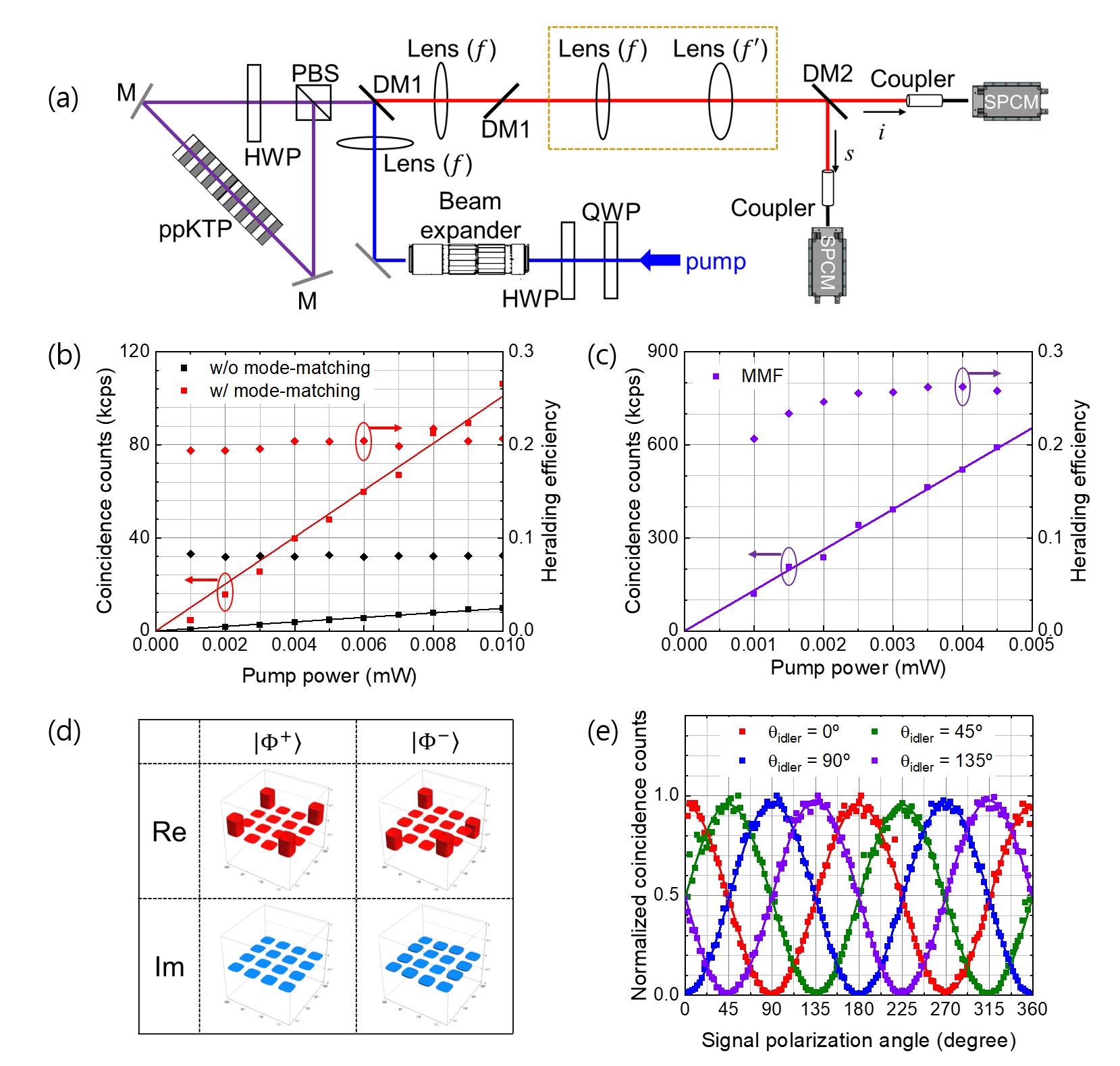}
\caption{(a) Ultrabright polarization-entangled photon source with mode-matching. (b) Coincidence counts (square) and heralding efficiencies (diamond) as a function of the pump power, comparing results with (red) and without (black) mode-maching. The slopes indicate linear fits of measured data. (c) Coincidence counts (square) and heralding efficiencies (diamond) as a function of the pump power collected by multimode optical fiber (MMF).The slope indicates linear fit of measured data. (d) Reconstructed density matrix for  $\vert\Phi^{\pm}\rangle$ states and the fidelities. (e) Polarization-correlation functions obtained for H/V and D/A measurement basis. The visibilities calculated from the fits to raw data are 0.976 (H), 0.986 (V), 0.971 (D), and 0.979 (A). The polarization-correlation data for each basis has been normalized to its maximum coincidence count value.}
\label{fig5}
\end{figure}
Figure \ref{fig4}(d) and \ref{fig4}(e) depict the data obtained by the coincidence measurements and heralding efficiencies as a function of the pump waist.
Signal and idler photons are coupled into SMFs, and single and coincidence events are counted by SPCM and TCSPC module.
Heralding efficiency is defined as $\eta_{h}=C_{c}/\sqrt{C_{s}C_{i}}=\sqrt{T_{s}T_{i}\eta_{s}\eta_{i}D_{s}D_{i}}$, where $C_{s,i,c}$ refer to signal, idler, and coincidence counts, respectively.
In the setup, $T_{s(i)}\sim0.65$ (0.70) is the channel transmission, $D_{s(i)}\sim0.55$ (0.51) is detector efficiency, and $\eta_{s(i)}$ is a SMF coupling efficiency depending on $w_{p}$ and $f'$.
The pump waist is varied by using a beam expander and focusing lens ($f=250$ mm) while keeping the collimation lens $f'$ fixed at 50, 75, and 100 mm.
The colored solid lines in Fig. \ref{fig4}(d) represent theoretical curves considering only collinear Gaussian modes, corresponding to $w_{s}=25$ (red), 45 (green), and 65 (blue) $\mu$m.
The colored solid lines in Fig. \ref{fig4}(e) represent theoretical curves of heralding efficiencies for different collimation conditions.
The parameter $\eta_{h}$ given by $\sqrt{T_{s}T_{i}\eta_{s}\eta_{i}D_{s}D_{i}}$ depends only on $\eta_{s(i)}$, as $T_{s(i)}$ and $D_{s(i)}$ are constant.
As shown in Fig. \ref{fig3}(b), $w_{s}$ varies over a wide range with different generation rates for a fixed $w_{p}$ and each $w_{s}$ component couples differently to the SMF.
If the mode size of photon-pairs at the lens ($f'$) is smaller than the coupling mode size, the focused size of photon-pairs will exceed the SMF core size.
If the pair size at the lens is larger, the focused size of pairs will be smaller than the SMF core size, but the incident angle may exceed the fiber's numerical aperture.

\begin{table*}[t]
\centering
\caption{State-of-art sources of polarization-entangled photons based on type-0 SPDC. Heralding efficiency ($\eta_{h}$) is defined as (i) $C_{c}/C_{s(i)}$ or (ii) $C_{c}/\sqrt{C_{s}C_{i}}$. $^{*}$ denotes values estimated from published data.}
\begin{tabular}{p{90pt} p{60pt} p{90pt} p{80pt} p{60pt} p{35pt}}
\hline
Reference &   $L$        & $\lambda_{s}$, $\lambda_{i}$ ($\Delta \lambda$) & $B=\frac{C_{c}}{\Delta\lambda\cdot P_{p}}$ & $\eta_{h}$ & Fidelity  \\
                     & (mm) &      (nm)                                                                 &                        (MHz/mW/nm)                          &                   & $(\%)$ \\
\hline
Hentschel et al.\cite{OE09Hentschel}  & 30 & 810,1550 (0.4,1.5)& 0.0062$^{*}$ & 0.016 $^{\text{i}}$ & 98.2 \\
\hline
Steinlechner et al.\cite{OE12Steinlechner} & 20 & 784,839 (2.3) & 0.28 & 0.18 $^{\text{i}}$ & 98.3 \\
\hline
Steinlechner et al.\cite{JOSAB14Steinlechner} & 20 & 784,839 (2.3)& 0.22 & 0.31 $^{\text{i}}$ & 99.1 \\
\hline
Jabir et al. \cite{SR17Jabir} & 30 & 810 (2) & 0.019 & 0.185 $^{\text{ii}}$ & 97.5 \\
\hline
Chen et al. \cite{PRL18Chen}  & 11.48 & 810 (3) & 0.053 & 0.185 $^{\text{i}}$ & 99.2 \\
\hline
Lohrmann et al. \cite{APL20Lohrmann} &  & 810 (14) & 0.1$^{*}$ & 0.22 $^{\text{i}}$ &  \\
\hline
Brambila et al. \cite{OE23Brambila} & 30 & 810 (0.8) & 1.07 & 0.2 & 99 \\
\hline
this work & 30 & 796,824 (4.9) & 2.06 & 0.2  $^{\text{ii}}$ & 97.8  \\
\hline
\end{tabular}
 \label{parameters}
\end{table*}

In this case, coupling is estimated by the ratio of the photon-pair mode size at the lens to the coupling mode size.
The overall coupling efficiency is the ratio of total generated photon-pairs to total coupled photon-pairs across all $w_{s}$ components for a fixed $w_{p}$.
Using a collimation lens with $f'=$ 50 mm for $w_{p}=$ 20 $\mu$m, $C_{s(i)}=847$ kcps (768 kcps) and $C_{c}=$ 214 kcps at a pump power $P_{p}=$ 10 $\mu$W.
This results in a coincidence counts rate of 21.4 MHz/mW, with $\eta_{h}=$ 0.265 and $\eta_{s}=\eta_{i}=0.742$ (assuming $\eta_{s}=\eta_{i}$).
The spectral bandwidth ($\Delta\lambda$) at FWHM for the signal and idler is 4.915 and 4.911 nm respectively (see Fig. \ref{fig2}(a)), leading to a spectral brightness, $B=\frac{C_{c}}{\Delta\lambda\cdot P_{p}}\sim$ 4.4 MHz/mW/nm.
For $f'=75$ mm and $w_{p}=30$ $\mu$m, $\eta_{s(i)}=$ 0.760, $C_{s(i)}=636$ kcps (527 kcps) and $C_{c}=$ 158 kcps, resulting in $\eta_{h}=$ 0.272 and $B\sim$ 3.2 MHz/mW/nm.
For $f'=100$ mm and $w_{p}=40$ $\mu$m, $\eta_{s(i)}=$ 0.626, $C_{s(i)}=477$ kcps (473 kcps), $C_{c}=$ 107 kcps, $\eta_{h}=$ 0.224, and $B\sim$ 2.2 MHz/mW/nm.
Despite the highest coupling efficiency at $f'=75$ mm, the coincidence count rate is highest under the condition of $f'=50$ mm.
This is because the photon-pair generation rate is significantly higher when the pump beam is more tightly focused to $w_{p}=20$ $\mu$m.

\section{Ultrabright polarization-entangled photon source}

In this section, we demonstrate the implementation of an ultrabright polarization-entangled photon source based on a Sagnac interferometer, building on the results previously presented.
Collinearly emitted photon-pairs, comprising signal ($\sim$796 nm) and idler ($\sim$824 nm), travel both clockwise and counter-clockwise within the Sagnac interferometer to form polarization-entangled states, represented as $\vert\Phi^{\pm}\rangle=\frac{1}{\sqrt{2}}\left( \vert HH \rangle \pm \vert VV \rangle \right)$.
Figure \ref{fig5}(a) depicts the setup of a highly-efficient fiber-coupled entangled photon source, achieved through mode-matching between the fundamental mode of the collection optics and the modes of the polarization-entangled photons.
The pump beam is focused by the lens with $f=$ 250 mm, and the polarization-entangled photons emerging from the interferometer are collimated again by a lens with $f=$ 250 mm before being adjusted to the optimal mode size through a combination of lenses with $f=$ 250 mm and $f'=$ 50 mm, and then coupled with SMFs.

When applying the mode-matching scheme, the single count $C_{s(i)}$ and coincidence count $C_{c}$ show 57.5 MHz/mW (41.7 MHz/mW) and 10.1 MHz/mW respectively, with a corresponding spectral brightness $B\sim$ 2.06 MHz/mW/nm as shown in Fig. \ref{fig5}(b).
This represents an over tenfold increase in spectral brightness compared to the case without mode-matching, and more than a twofold improvement in the heralding efficiency, increasing from 0.08 to 0.2.
The results obtained in this work exhibit the highest value among the reported high-efficiency polarization-entangled photon sources to date.
Table 1 summarizes the parameter values for each reference.

Additionally, coincidence counts measured using a MMF and their corresponding heralding efficiencies are shown in Fig. \ref{fig5}(c).
MMFs, due to their inherently large core size, capture the majority of collinear entangled photon-pairs.
As a result, high coincidence counts of 131 MHz/mW and a heralding efficiency of 0.25 are achieved.

To assess the degree of entanglement a quantum state tomography \cite{PRA01James} are performed, whereby a Bell state fidelities $F=\langle\Phi^{\pm}\vert\rho\vert\Phi^{\pm}\rangle$ of 0.978 $\pm$ 0.0049 and 0.977 $\pm$ 0.0049 respectively at the pump power of 0.01 mW (Fig. \ref{fig5}(d)).
The polarization-entanglement is also characterized by the polarization-correlation measurements in two mutually unbiased bases (Fig. \ref{fig5}(e)).
The Bell-CHSH value $S$ is 2.782 $\pm$ 0.04 \cite{PRL69Clauser}, where the error is estimated from the shot noise assuming a Poisson distribution of photons.

\section{Conclusion}
In this study, we implement a high-efficiency polarization-entangled photon source based on SMF coupling with a spectral brightness exceeding 2.0 MHz/mW/nm, utilizing a spatial mode-matching technique.
The entangled photon-pairs generated by the nonlinear interaction between the crystal and the pump beam have their spectra determined by the crystal's length and temperature, and the beam waist distribution of the signal and idler is influenced by the pump waist focused on the crystal.
We carefully examine the characteristics of SPDC photons under varying experimental conditions through both experimental and theoretical approaches.
Especially, by matching the collinear Gaussian modes of SPDC photons with the fundamental mode of the collection optics, we identify the optimal conditions for extracting the maximum number of entangled photon-pairs.
The results of this study provide a more intuitive and practical guide for implementing high-efficiency quantum light sources necessary for long-distance free-space quantum communication, where channel loss is extremely severe.

\begin{acknowledgments}
We thank Prof. Heonoh Kim for helpful comments.
This work was supported by Agency for Defense Development.
\end{acknowledgments}


\begin{thebibliography}{99}

\bibitem{Nature08Kimble} H. J. Kimble, ``The quantum internet'', \textit{Nature} \textbf{453}, 1023 (2008).

\bibitem{NP16Vlaivarthi} R. Valivarthi, M. G. Puigibert, Q. Zhou, G. H. Aguilar, V. B. Verma, F. Marsili, M. D. Shaw, S. W. Nam, D. Oblak, and W. Tittel, ``Quantum teleportation across a metropolitan fibre network", \textit{Nat. Photonics} \textbf{10}, 676 (2016).

\bibitem{Science17Yin} J. Yin, et al., ``Satellite-based entanglement distribution over 1200 kilometers",  \textit{Science} {\bf 356}, 1140 (2017).

\bibitem{Nature17Liao} S.-K. Liao, at al., ``Satellite-to-ground quantum key distribution",  \textit{Nature} \textbf{549}, 43 (2017).

\bibitem{RMP22Lu} C. Y. Lu, Y. Cao, C. Z. Peng, and J. W. Pan, ``Micius quantum experiments in space", \textit{Rev. Mod. Phys.} \textbf{94(3)}, 035001(2022).

\bibitem{PRL18Cao} Y. Cao, Y.-H. Li, W.-J. Zou, Z.-P. Li, Q. Shen, S.-K. Liao, J.-G. Ren, J. Yin, Y.-A. Chen, C.-Z. Peng, and J.-W. Pan, ``Bell test over extremely high-loss channels: Towards distributing entangled photon pairs between Earth and the Moon", \textit{Phys. Rev. Lett.} \textbf{120}, 140405 (2018).

\bibitem{ACSphoton21Zeuner} K. D. Zeuner, K. D. J$\ddot{\text{o}}$ns, L. Schweickert, C. R. Hedlund, C. N. Lobato, T. Lettner, K. Wang, S.Gyger, E. Sch$\ddot{\text{o}}$ll, S. Steinhauer, M. Hammar, and V. Zwiller, ``On-demand generation of entangled photon pairs in the telecom C-band with InAs quantum dots",  \textit{ACS Photonics} \textbf{8}, 2337 (2021).

\bibitem{PRXquantum21Steiner} T. J. Steiner, J. E. Castro, L. Chang, Q. Dang, W. Xie, J. Norman, J. E. Bowers, and G. Moody, ``Ultrabright entangled-photon-pair generation from an AlGaAs-on-insulator microring resonator", \textit{PRX Quantum} \textbf{2}, 010337 (2021).

\bibitem{QST23Bruns} A. Bruns, C.-Y. Hsu, S. Stryzhenko, E. Giese, L. P. Yatsenko, I. A. Yu, T. Halfmann, and T. Peters, ``Ultrabright and narrowband intra-fiber biphoton source at ultralow pump power", \textit{Quantum Sci. Technol.} {\bf 8}, 015002 (2023).

\bibitem{arXiv24Shiu} J.-S. Shiu, Z.-Y. Liu, C.-Y. Cheng, Y.-C. Huang, I. A. Yu, Y.-C. Chen, C.-S. Chuu, C.-M. Li, S.-Y. Wang, and Y.-F. Chen, ``Observation of highly correlated ultrabright through increased atomic ensemble density in spontaneous four-wave mixing", arXiv:2312.12758v3 (2024).

\bibitem{PRL88Shih} Y. H. Shih and C. O. Alley, ``New type of Einstein-Podolsky-Rosen-Bohm experiment using pairs of light quanta produced by optical parametric down conversion", \textit{Phys. Rev. Lett.} \textbf{61}, 2921 (1988).

\bibitem{PRL95Kwiat} P. G. Kwiat, K. Mattle, H. Weinfurter, A. Zeilinger, A. V. Sergienko, and Y. Shih, ``New high-intensity source of polarization-entangled photon pairs", \textit{Phys. Rev. Lett.} \textbf{75}, 4337 (1995).

\bibitem{OE07Fujii} G. Fujii, N. Namekata, M. Motoya, S. Kurimura, and S. Inoue, ``Bright narrowband source of photon-pairs at optical telecommunication wavelength using type-II periodically poled lithium niobate waveguide", \textit{Opt. Express} {\bf 15}, 12769 (2007).

\bibitem{OE09Chen} J. Chen, A. J. Pearlman, A. Ling, J. Fan, and A. Migdall, ``A versatile waveguide source of photon pairs for chip-scale quantum information processing", \textit{Opt. Express} {\bf 17}, 6727 (2009).

\bibitem{OE12Steinlechner} F. Steinlechner, P. Trojek, M. Jofre, H. Weier, D. Perez, T. Jennewein, R. Ursin, J. Rarity, M. W. Mitchell, J. P. Torres, H. Weinfurter, and V. Pruneri, ``A high-brightness source of polarization-entangled photons optimized for application in free space", \textit{Opt. Express} {\bf 20}, 9640 (2012).

\bibitem{JOSAB14Steinlechner} F. Steinlechner, M. Gilaberte, M. Jofre, T. Scheidl, J. P. Torres, V. Pruneri, and R. Ursin, ``Efficient heralding of polarization-entangled photons from type-0 and type-II spontaneous parametric downconversion in periodically poled KTiOPO$_4$," \textit{J. Opt. Soc. Am. B} {\bf 31}, 2068 (2014).

\bibitem{SR17Jabir} M. V. Javir and G. K. Samanta, ``Robust, high brightness, degenerate entangled photon source at room temperature," \textit{Sci. Rep.} {\bf 7}, 12613 (2017).

\bibitem{OE08Fiorentino} M. Fiorentino and R. G. Beausoleil, ``Compact source of polarization-entangled photons", \textit{Opt. Express} {\bf 16}, 20149 (2008).

\bibitem{PRA05Lee} P. S. K. Lee, M. P. van Exter, and J. P. Woerdman, ``How focused pumping affects type-II spontaneous parametric down-conversion", \textit{Phys. Rev. A} {\bf 72}, 033803 (2005).

\bibitem{OE11Halevy} A. Halevy, E. Megidish, L. Dovrat, H. S. Eisenberg, P. Becker, and L. Bohaty, ``The biaxial nonlinear crystal BiB$_3$O$_6$ as a polarization entangled photon source using non-collinear type-II parametric down-conversion", \textit{Opt. Express} {\bf 19}, 20420 (2011).

\bibitem{OE13Jeong} Y.-C. Jeong, K.-H. Hong, and Y. H. Kim, ``Bright source of polarization-entangled photons using a PPKTP pumped by a broadband multi-mode diode laser", \textit{Opt. Express} {\bf 24}, 1165 (2016).

\bibitem{OE13Steinlechner} F. Steinlechner, S. Ramelow, M. Jofre, M. Gilaberte, T. Jennewein, J. P. Torres, M. W. Mitchell, and V. Pruneri ``Phase-stable source of polarization-entangled photons in a linear double-pass configuration", \textit{Opt. Express} {\bf 21}, 11943 (2013).

\bibitem{APL20Lohrmann} A. Lohrmann, C. Perumangatt, A. Villar, and A. Ling, ``Broadband pumped polarization entangled photon-pair source in a linear beam displacement interferometer", \textit{Appl. Phys. Lett.} {\bf 116}, 021101 (2020).

\bibitem{QST21Lee} Y. S. Lee, M. Xie, R. Tannous, and T. Jennewein, ``Sagnac-type entangled photon source using only conventional polarization optics", \textit{Quantum Sci. Technol.} {\bf 6}, 025004 (2021).

\bibitem{PRA05Ljunggren} D. Ljunggren and M. Tengner, ``Optimal focusing for maximal collection of entangled narrow-band photon pairs into single-mode fibers", \textit{Phys. Rev. A} {\bf 72}, 062301 (2005).

\bibitem{PRA10Bennink} R. S. Bennink, ``Optical collinear Gaussian beams for spontaneous parametric down-conversion," \textit{Phys. Rev. A} {\bf 81}, 053805 (2010).

\bibitem{OE07Fedrizzi} A. Fedrizzi, T. Herbst, A. Poppe, T. Jennewein, and A. Zeilinger, ``A wavelength-tunable fiber-coupled source of narrowband entangled photons," \textit{Opt. Express} \textbf{15}, 15377 (2007).

\bibitem{OE13Guerreiro} T. Guerreiro, A. Martin, B. Sanguinetti, N. Bruno, H. Zbinden, and R. T. Thew, ``High efficiency coupling of photon pairs in practice," \textit{Opt. Express} \textbf{21}, 27641 (2013).

\bibitem{CRSSM99Satyanarayn} M. N. Satyanarayan, A. Deepthy, and H. L. Bhat, ``Potassium titanyl phosphate and its isomorphs: Growth, properties, and application", \textit{Crit. Rev. Solid State Mater. Sci.}, \textbf{24}, 103 (1999).

\bibitem{PRA01James} D. F. V. James, P. G. Kwiat, W. J. Munro, and A. G. White, ``Measurement of qubits", \textit{Phys. Rev. A}, \textbf{64}, 052312 (2001).

\bibitem{PRL69Clauser} J. F. Clauser, M. A. Horne, A. Shimony, and R. A. Holt, ``Proposed experiment to test local hidden variable theories," \textit{Phys. Rev. Lett.} \textbf{23}, 880 (1969).

\bibitem{OE09Hentschel} M. Hentschel, H. H$\ddot{\text{u}}$bel, A. Poppe, and A. Zeilinger, ``Three-color Sagnac source of polarization-entangled photon pairs," \textit{Opt. Express} \textbf{17}, 23153 (2009).

\bibitem{PRL18Chen} Y. Chen, S. Ecker, S. Wengerowsky, L. Bulla, S. K. Joshi, F. Steinlechner, and R. Ursin, ``Polarization entanglement by time-reversed Hong-Ou-Mandel interference," \textit{Phys. Rev. Lett.} \textbf{121}, 200502 (2018).

\bibitem{OE23Brambila} E. Brambila, R. G$\acute{\text{o}}$mez, R. Fazili, M. Gr$\ddot{\text{a}}$fe, and F. Steinlechner, ``Ultrabright polarization-entangled photon-pair source for frequency-multiplexed quantum communication in free-space," \textit{Opt. Express} \textbf{31}, 16107 (2023).

\end{thebibliography}
\end{document}